\documentclass[12pt]{iopart}
\usepackage{iopams}  
\usepackage{xcolor}

\expandafter\let\csname equation*\endcsname\relax
\expandafter\let\csname endequation*\endcsname\relax
\usepackage{physics}

\begin{document}

\title{On the Hydrogen Atom in the Holographic Universe}

\author{S. Jalalzadeh$^1$,  S. Abarghouei Nejad$^1$ and P.V. Moniz$^{2,3}$}
\address{$^1$Departamento de F\'{i}sica, Universidade Federal de Pernambuco,
Recife, PE, 52171-900, Brazil}
\address{$^2$Departmento de F\'{i}sica , Universidade da Beira Interior, 6200 Covilh\~a, Portugal}
\address{$^3$Centro de Matem\'atica e Aplica\c c\~oes (CMA-UBI), Covilh\~a, Portugal}
\ead{shahram.jalalzadeh@ufpe.br}
\ead{salmanabar@df.ufpe.br}
\ead{pmoniz@ubi.pt}
\vspace{10pt}

\begin{abstract}
We investigate the holographic bound utilizing a homogeneous, isotropic, and non-relativistic neutral hydrogen gas present in the de Sitter space. Concretely, we propose to employ de Sitter holography intertwined with quantum deformation of the hydrogen atom using the framework of quantum groups.
Particularly, the $\mathcal U_q(so(4))$ quantum algebra is used to construct a finite-dimensional Hilbert space of the hydrogen atom. As a consequence of the quantum deformation of the hydrogen atom, we demonstrate that the Rydberg constant is dependent on the de Sitter radius, $L_\Lambda$. This feature is then extended to obtain a finite-dimensional  Hilbert space for the full set of all hydrogen atoms in the de Sitter universe. We then show that the dimension of the latter Hilbert space satisfies the holographic bound. We further show that the mass of a hydrogen atom $m_\text{atom}$, the total number of hydrogen atoms at the universe, $N$, and the retrieved dimension of the Hilbert space of neutral hydrogen gas, $\text{Dim}{\mathcal H}_\text{bulk}$, are related to the de Sitter entropy, $S_\text{dS}$, the Planck mass, $m_\text{Planck}$, the electron mass, $m_\text{e}$, and the proton mass $m_\text{p}$, by $m_\text{atom}\simeq m_\text{Planck}S_\text{dS}^{-\frac{1}{6}}$, $N\simeq S_\text{dS}^\frac{2}{3}$ and $\text{Dim}{\mathcal H}_\text{bulk}=2^{\frac{m_\text{e}}{m_\text{p}}\alpha^2S_\text{dS}}$, respectively.
\end{abstract}
\vspace{2pc}
\submitto{\PS}
\maketitle

\section{Introduction}

\indent 

The seminal result establishing a distinctive bound on the entropy of  a spacelike region of spacetime was formulated initially by Bekenstein \cite{bekenstein1981universal}, and subsequently, it was 
gradually advanced upon fundamental research on  the thermodynamic
properties of massive black holes and other gravitational settings alike \cite{bekenstein1973black,bekenstein1974generalized,hawking1975particle}. More recently, it was broadly 
extended into what is known as Bousso's 
covariant 
entropy conjecture \cite{bousso1999covariant}, which 
conveys a well-defined holographic bound. In essence, 
it claims to be to be a feature of any physical theory in that $S\leq \frac{A}{4G}$, where $G$ is the
Newton's gravitational constant and $A$ is the area
bounding any region,  which satisfies the spacelike projection theorem\footnote{It may be worthy of stressing the difference between the ``holographic bound" and what is more appropriately known as ``Bekenstein bound," i.e., $S<ER$, where $E$ is the energy contained in a region of size $R$). Generally, $S<ER<A/4G$ \cite{2020arXiv200912530B,10.2307/26060403}.} \cite{bousso1999covariant}. These ideas were build from  't Hooft \cite{tHooft:1993dmi},  {Fischler \cite{fischler1998holography}} and Susskind \cite{susskind1995world}
 proposal of  the holographic principle.

Let us be more concrete and start by summarizing the textbook thermodynamic Bekenstein bound for black holes plus its weak and strong forms, respectively, adopting from 
Smolin \cite{smolin2001strong} as follows:
\begin{itemize}
    \item The thermodynamic black hole entropy is  $S_{bh}=A_{bh}/4G$, where $A_{bh}$ is the area of the black hole horizon, constituting a crucial feature of the laws of black hole mechanics.
    \item The weak black hole entropy a measure of how much information can be gathered by an external observer about the interior of the black hole from measurements made outside the horizon. Besides the mass, angular momentum, and charge of the black hole, the aforementioned includes measurements of the radiation emitted by the black hole.
    \item The strong black hole entropy is  a measure of how much information or number of degrees of freedom are encompassed in the interior region of the black hole. 
\end{itemize}

These can be generalized for a bulk space $\mathcal V_\text{bulk}$ with a fixed boundary $\Sigma=\partial\mathcal V_\text{bulk}$   \cite{smolin2001strong}, namely: 
\begin{enumerate}
    \item Weak holographic 
bound: Let  ${\mathcal H}_{\Sigma}$
    be the Hilbert  space of observables on the boundary $\Sigma$. Then
    \begin{eqnarray}
    \text{Dim}~{\mathcal H}_\Sigma\leq e^{\frac{A_\Sigma}{4G}},
    \end{eqnarray}
    where $A_\Sigma$ is the area of $\Sigma$ and $\text{Dim}~{\mathcal H}_\Sigma$ is the smallest appropriate Hilbert space ${\mathcal H}_\Sigma$. 
        \item  
        Holographic bound:  Let $\mathcal H_\text{bulk}$ be the smallest Hilbert space of local observables measurable in the interior
of a volume with boundary $A$  in the bulk space. Then
        \begin{eqnarray}\label{Smolin}
\text{Dim}\,\, \mathcal H_\text{bulk}\leq \exp\left({\frac{A}{4G}}\right),
\end{eqnarray}
where $\text{Dim}\,\, \mathcal H_\text{bulk}$ denotes
the dimension of the Hilbert space. This latter pronouncement constitutes the core and essence of  Bousso's covariant entropy conjecture \cite{bousso1999covariant}, 
establishing a precise holographic bound 
for physical theories. It will be about this one that we will elaborate our paper. 
\end{enumerate}

In addition, the above described holographic bound  was widely applied  on cosmological
solutions and  other gravitational collapsing systems.
It has been studied employing several spacetimes,  enlarging the diversity of its  appraisal, by means of a significant set of contributions to  the literature \cite{bousso2004bound,casini2008relative,bousso2002holographic,bousso2000positive,2008PhRvD..78f4047A,2000CQGra..17L..83B}, all confirming it. 
As remarked herewith, it was  further advanced by Bousso in  that it may be a universal law of nature \cite{bousso1999covariant,bousso1999holography},
 within a background-independent formulation and bearing a holographic description of nature  at its inception.  Notwithstanding its allure,  the conjecture presented 
 in \cite{bousso1999covariant,bousso1999holography}, albeit quite successful,  has not yet been proven. 
 Approaches and surveys to find a route  to do so,  have so far involved   general relativity or standard field theory \cite{bousso1999covariant,bousso1999holography,bousso2004bound,casini2008relative,bousso2002holographic,bousso2000positive}.

Within the context conveyed by the above paragraphs, the unpretentious purpose in our manuscript is to introduce a 
  discussion on the holographic bound from another angle. We are not saying the conjecture cannot be proven strictly in the terms advocated throughout the herewith mentioned references \cite{bousso2004bound,casini2008relative,bousso2002holographic,bousso2000positive}; we are merely proposing to bring to  the discussion a different (hopefully complementary) perspective, broadening the scope of 
discussion by means of another framework and tools. 

Being more specific, our  work will look at it but within a more  basic and less  complicated way. Concretely,   employing  a twofold research that constitutes the main contribution of this paper. 
 On the one hand,  it is argued that the existence of the de Sitter (DeS) horizon, which satisfies the spacelike projection theorem,  suggests a holographic realization for the hydrogen atom gas. In particular, the infinite-dimensional Hilbert space of the bound states of an atom is  inconsistent with the holographic principle. This motivates us to introduce and explore the features of quantum deformation (within quantum groups) as a tool to bring more evidence supporting the universality of the holographic principle and the corresponding 
 bound. A concrete quantum algebra is used to construct a finite dimensional Hilbert space of the hydrogen atom whose  Rydberg constant is then shown to be dependent on the DeS radius.  On the other hand, we 
subsequently take a $(3+1)$-dimensional spacetime filled with non-relativistic matter. A finite-dimensional  Hilbert space for the  set of   hydrogen atoms in
the DeS universe is estimated. We then show that the dimension of the corresponding  Hilbert space satisfies the holographic bound plus that several quantities become consequently intertwined within the dimensions of the Hilbert space and the DeS entropy.

\section{Quantum deformation of the hydrogen atom} 

\indent

The current observational paradigm presents our universe as accelerating and the cosmic event horizon {increases monotonically, asymptotic} to a {specific}  radius. 
{Hence}, it is {reasonable to employ the working assumption where the}  late time universe is a DeS space,   with a cosmic event horizon equal to the DeS radius $L_\Lambda:=\sqrt{\frac{3}{\Lambda}}= 16.4\pm0.4\,\,\text{ Glyr} = (1.55\pm0.04)\times10^{26}\,\text{ m}$. So,
let us {assume} a non-relativistic hydrogen atom located at the origin of {a DeS}
space with local coordinates $(t,r,\theta,\phi)$. 

{ The cosmological constant is very small and that implies 
the observable universe to be large and nearly flat, so we will consider a usual non-relativistic Hamiltonian operator of the hydrogen atom in which the fine structure,  all quantum field corrections are considered as a perturbation to it. In addition, we assume that the gravitational correction of the spacetime curvature to the Schr\"odinger equation \cite{moradi2010hydrogen,parker1980one,parker1982gravitational}
\begin{eqnarray}
V_\text{g}=\frac{m_\text{e}}{2}R_{oioj}x^ix^j,
\end{eqnarray}
{constitutes} a small perturbation potential, with $R_{0i0j}$ {being} the Riemann tensor in Fermi normal coordinates (where the metric is rectangular and has vanishing first derivatives at each point of a curve),  $x^i$ {is} the position of the electron in the nucleus-centered and $m_\text{e}$ is the electron mass. 
The {subsequent} spectrum of the hydrogen atom at such spacetime is
\begin{eqnarray}\label{Energy}
E_n=-\frac{\alpha^2}{2m_\textbf{e}n^2}+\frac{A_{n,j}}{4\alpha^2m_\text{e}L_\Lambda^2}+{\mathcal O}\left(\frac{1}{n^3}\right),
\end{eqnarray}
where the second term in the above expression of the energy is the energy shift of the non-relativistic correction regarding the presence of the cosmological constant \cite{moradi2010hydrogen}, $A_{nj}$ are  constants dependent to the quantum numbers of the state and ${\mathcal O}(1/n^3)$ represents the fine structure, the hyperfine structure and other corrections from quantum field theory such as the Lamb shift and the anomalous magnetic dipole moment of the electron.  
Hence, the wavelength of an emitted photon by a hydrogen atom, $\lambda$, is given by the modified Rydberg formula 
\begin{eqnarray}\label{1aaa}
\frac{1}{\lambda}=\frac{m_\text{e}\alpha^2}{4\pi}\left(\frac{1}{n_f^2}-\frac{1}{n_i^2}\right)+\frac{1}{4m_\text{e}\alpha^2L_\Lambda^2}(A_{n_i,j_i}-A_{n_f,j_f})+{\mathcal O}(\frac{1}{n^3}),
\end{eqnarray}
where $\alpha$ is the fine structure constant and $n_i,n_f$ are the principal quantum numbers of initial and final states involved in the transition, respectively.} 
 According to the Rydberg  expression for the hydrogen atom spectra (\ref{1aaa}), 
 the wavelength of an emitted photon between two successive states $n_i=n$ and $n_f=n-1$, and $n\gg1$ is given by $\lambda\simeq\frac{2\pi}{m_\text{e}\alpha^2}n^3$ for arbitrarily large values of $n$. {Note that in obtaining this relation, by assuming $n\gg1$ and $1/L_\Lambda^2\ll1$, we kept only the first term of the modified Rydberg formula.}

 Additionally, allow us to mention that an infinite-dimensional Hilbert space 
 for an atom conveys an  inconsistency  with the holographic bound (\ref{Smolin}). 
 This argument can be further elaborated as follows. 
 In the presence of a cosmological constant
$\Lambda$, any local observer eventually perceives the space as a box of size $L_\Lambda$. Therefore, in the presence of a cosmological horizon, as far as bound states are concerned, all adjacent transitions with $n_i\geq \left(\frac{\alpha^2L_\Lambda m_\text{e}}{2\pi }\right)^\frac{1}{3}$ are forbidden. 
To clarify,  let us consider a generic transition between two states $n_f$ and $n_i$ in which $n_i=n_f+k$ (in order to include adjacent ($k=1$) as well as others); the wavelength of the emitted photon, $\lambda$, must be less than the radius of the horizon $L_\Lambda$. Using the Rydberg formula, for large values of $n_i$ and for a given $k$, $n_i$ 
must satisfy the following inequality:
$\frac{2\pi}{\alpha^2m_\text{e}}\frac{n_i^3}{k}\leq L_\Lambda$.
For example, with $k=10$   all transitions  $n_i\geq\left(\frac{5\alpha^2L_\Lambda m_\text{e}}{\pi }\right)^\frac{1}{3}$ are forbidden. 
Moreover, for transitions with larger $k$, subsequent larger values of $n_i$  will be forbidden. 
Thus, there is a restriction on the maximal wavelength of the 
emitted photon, namely  $\lambda\leq L_\Lambda$. Saturation of this inequality gives the highest allowed principal quantum number, namely 
\begin{eqnarray}\label{0-1}
n^3_\text{max}=\frac{\alpha^2m_\text{e}}{2\pi}L_\Lambda
\end{eqnarray}
in DeS space. Hence, the existence of 
the DeS horizon, which satisfies the spacelike projection theorem,  suggests a holographic realization for the hydrogen atom, whose Hilbert space ought to be
 finite-dimensional.

 In the context conveyed throughout the  previous paragraph,
 the purpose of the  herein paper is to 
 analyse whether a degree of holography can be 
 suitably brought 
 to discuss other features, namely the 
 holographic entropy bound.
 Standard quantization methods,  adopting (\ref{0-1}), would 
  merely constrain $n_\text{max}$. This obvious result, its interest notwithstanding,  can be surpassed
 if we instead employ the features of quantum deformations
 (within quantum groups) as  a tool.

There are a number of ways to construct a finite-dimensional Hilbert space. One method to {retrieve the dimension of an Hilbert space into a finite value is through quantum deformation (by means of quantum groups)
of the model, when the deformation parameter is a root of unity} \cite{chaichian1996introduction}. { Deformed Hydrogen atom models are studied in different ways, such as using moyal-like noncommutative product as \cite{kupriyanov2013hydrogen} and \cite{chaichian2004non}, or Kustaanheimo Stiefel transformation \cite{Kustaanheimo:419610,boiteux1973three}. Here, we study a $ q$-deformatrion of dynamical symmetry of Hydrogen atom by using the quantum group $ so_q(4) $. This is done (as in the case of a real deformation parameter, $q\in\mathbb R$, used in Refs. \cite{boiteux1973three,kibler1991q,song1992quantum,gora1992two,yang1993energy,castro2015physics}) by enlarging the corresponding symmetry group, using the Laplace-Runge-Lenz vector, and the separation of $ so_q(4) = su_q(2) \otimes su_q(2) $.}

Historically, quantum groups {have emerged from  studies on quantum integrable models, using quantum inverse scattering
methods,  which led to deformation of classical matrix groups and their corresponding Lie algebras} \cite{kulish1983quantum,sklyanin1982some,faddeev2016liouville}. Recently, quantum groups {were found to} play a major role in quantum integrable systems \cite{chari1995guide},
conformal field theory \cite{oh1992conformal}, knot theory \cite{kauffman2007q}, solvable lattice models \cite{foda1994vertex}, topological quantum computations \cite{nayak2008non}, 
molecular spectroscopy \cite{chang1991q} and quantum gravity \cite{smolin1995linking,major1996quantum,dupuis2013quantum,dittrich2014quantum,rovelli2015compact}.

\subsection{Standard tools}
In our paper, we shall deal with the quantum deformation of the universal  enveloping algebra,
$so(4)\cong su(2)\otimes su(2)$, of the hydrogen atom. Since we are interested in
$\mathcal{U}_q(su(2)\otimes  su(2))$, let us review first some basic facts
about the quantum group $\mathcal{U}_q(su(2))$. Explicitely,  $\mathcal{U}_q(su(2))$ is
a Hopf algebra over $\mathbb C$ generated by set of operators $\{q^{J_0},q^{-J_0},J_\pm\}$
satisfying relations { \cite{biedenharn1995quantum}}
\begin{eqnarray}\label{2}
q^{J_0}J_\pm q^{-J_0}=q^\pm J_\pm,\,\,\,\,
[J_+,J_-]=[2J_0]_q
\end{eqnarray}
where $q:=\exp(\frac{2\pi i}{\mathfrak{D}})$, is the deformation parameter with $\mathfrak{D}:=n_\text{max}\in\mathbb
N$, $\mathfrak{D}\geq2$, and
\begin{eqnarray}
[x]_q:=\frac{q^x-q^{-x}}{q-q^{-1}}=\frac{\sin(\frac{2\pi x}{\mathfrak{D}})}{\sin(\frac{2\pi}{\mathfrak{D}})}.
\end{eqnarray}
{For $\mathfrak D\rightarrow\infty$ (or equivalently using the definition of $\mathfrak D= n_{\text{max}} \simeq 
L_{\Lambda}^{\frac{1}{3}} \sim \Lambda^{-\frac{1}{6}}$, $\Lambda\rightarrow0$)}, $q\rightarrow1$ and we recover the Lie algebra
of $SU(2)$. The Casimir operator is given by
\begin{eqnarray}\label{3}
J^2:=\frac{1}{2}(J_+J_-+J_-J_+)+\frac{[2]_q}{2}[J_0]^2_q.
\end{eqnarray}
Let $\mathcal V_j=\{|jm\rangle,\,m=-j(1)j\}$ be the Hilbert space of the representation
theory of the ${\mathcal U}_q(su(2))$. Then 
\begin{eqnarray}\label{5}
J_\pm|jm\rangle=\sqrt{[j\pm m+1]_q[j\mp m]_q}|j,m\pm1\rangle,\cr
J_0|jm\rangle=m|jm\rangle,\cr
J^2|jm\rangle=[j]_q[j+1]_q|jm\rangle.
\end{eqnarray}
The invariants of $\mathcal{U}_q(su(2))$ at the root of unity
are $\{J^2,J_\pm^{\mathfrak{D}'},q^{\pm \mathfrak{D}'J_0}\}$ where 
\begin{eqnarray}\label{NN}
\mathfrak{D}':=\begin{cases}
\mathfrak{D}/2,\,\,\,\text{for even values of}\,\,\mathfrak{D},\\
(\mathfrak{D}-1)/2,\,\,\,\,\text{for odd values of}\,\,\mathfrak{D}. 
\end{cases}
\end{eqnarray}
At the nilpotent representation, which we are interested, $J_+^{\mathfrak{D}'}$ and
$J_-^{\mathfrak{D}'}$ have the zero eigenvalue for all eigenvectors, $J_\pm^{\mathfrak{D}'}|jm\rangle=0$.

Let us return to the bound states of a hydrogen atom. It is well known that the Hamiltonian
of the hydrogen atom
\begin{eqnarray}\label{7}
H=\frac{p^2}{2m_\text{e}}-\frac{\alpha}{r},
\end{eqnarray}
commutes with the orbital angular momentum $\mathbf{L}$ and the Laplace-Runge-Lenz vector $\mathbf M$. 
 { Namely, $ \textbf{M} = \frac{1}{2m_\text{e}}(\textbf{P} \times \textbf{L}-\textbf{L} \times \textbf{P}) - (\frac{\alpha\textbf{r}}{r}) $ . Furthermore, $H$, $\mathbf L$ and $\mathbf M$ satisfy the following relations \cite{adams2012algebraic}
\begin{eqnarray}\label{8}
\mathbf L.\mathbf M=0,\,\,\,\,\,\,\mathbf M^2=\alpha^2+\frac{2}{m_\text{e}}(\mathbf L^2+1)H,
\end{eqnarray}
 including following algebra,
\begin{eqnarray}
&& [L_i, H]= [M_i, H] = 0, \qquad [L_i,L_j] = i  \epsilon_{ijk} L_k, \cr
&& [L_i,M_j] = i  \epsilon_{ijk} M_k, ~~~ [M_i,M_j] = -\frac{2i }{m_\text{e}} \epsilon_{ijk} L_k H. 
\end{eqnarray}}
If we restrict ourselves to the bound states with energy $E$ and replace $H$ by $E$, then the vector operators $\mathbf L$ and $\tilde{\mathbf M}:=\mathbf
M\sqrt{-\frac{m_\text{e}}{2E}}$ satisfy the $so(4)$ commutation relations \cite{adams2012algebraic}.
 If we define the two vector operators 
\begin{eqnarray}\label{9}
\mathbf J^{(1)}:=\frac{1}{2}(\mathbf L+\tilde{\mathbf M}),\,\,\,\,\,\mathbf J^{(2)}:=\frac{1}{2}(\mathbf L-\tilde{\mathbf M}),
\end{eqnarray}
then the components of ${\mathbf J}^{(1)}$ and $\mathbf J^{(2)}$ satisfy the commutation relations of two commuting sets of $su(2)$ Lie algebras
\begin{eqnarray}\label{10}
[J^{(i)}_0,J^{(i)}_\pm]=\pm J^{(i)}_\pm,\,\,\,\,\,\,\,[J^{(i)}_+,J^{(i)}_-]=2J^{(i)}_0.
\end{eqnarray}
Note that these two sets of generators are not independents and we have the following two identities between Casimirs of $(su(2))_{(1)}$ and $(su(2))_{(2)}$
\cite{pauli1926wasserstoffspektrum}
\begin{eqnarray}\label{11}
\mathcal C_1:=\mathbf J_{(1)}^2-\mathbf J_{(2)}^2=0,\cr
\mathcal C_2:=\mathbf J_{(1)}^2+\mathbf J_{(2)}^2=-\frac{m_e\alpha^2}{4E}-\frac{1}{2}.
\end{eqnarray}
In fact, $\mathcal C_1$ and $\mathcal C_2$ are two independent Casimir operators of the original $SO(4)$ Lie group, in which $\mathcal C_1$ represents the
orthogonality
of the orbital angular momentum $\mathbf L$ and the Laplace-Runge-Lenz vector $\mathbf M$, and $\mathcal C_2$ is the Hamiltonian of the atom.
If we let $|j_1m_1j_2m_2\rangle$ denote the basis vectors for the $(su(2))_{(1)}\otimes
(su(2))_{(2)}$ the first Casimir in (\ref{11}) implies $j_1=j_2$ and the second Casimir
gives us the Bohr formula
\begin{eqnarray}\label{12}
E_n=-\frac{m_\text{e}\alpha^2}{2n^2},
\end{eqnarray}
where we identify $2j_1+1:=n$ as the principal quantum number.

\subsection{{ Quantum deformation and Hilbert space}}

One feasible way to define a
$q-$deformed hydrogen atom is to quantum deform the Lie groups  $(SU(2))_{(1)}$ and $(SU(2))_{(2)}$ each  of them defined by (\ref{2}). Also, the Pauli equations (\ref{11}) have to be extended to the quantum algebra $\mathcal U_q(su(2))_{(1)}\otimes 
\mathcal U_q(su(2))_{(2)}$. Then, this deformation via Eqs.(\ref{3}),
(\ref{5}) and (\ref{11}) produces a quantum deformed hydrogen atom with modified
Bohr formula given by
\begin{eqnarray}\label{13}
E_n=\frac{E_0}{4[j_1]_q[j_1+1]_q+1}=
\frac{E_0}{1+\frac{2}{\sin^2\left(\frac{2\pi}{\mathfrak{D}}\right)}\left(\cos(\frac{2\pi}{\mathfrak{D}})-\cos(\frac{2\pi
n}{\mathfrak{D}})\right)},
\end{eqnarray}
where $E_0:=-\frac{m_\text{e}\alpha^2}{2}$ is the energy of the ground state and
as the undeformed case, $n:=2j_1+1$ is the principal quantum number. Furthermore, the discrete spectrum exhibits the same degeneracy as that of the hydrogen atom in flat space.  It is clear that the $q-$deformed spectrum reduces to
that of the ordinary hydrogen atom when 
$\Lambda$ goes to zero.
{For  large values of $\mathfrak D$ and $n\ll\mathfrak D$, the expression of the 
$q-$deformed energy levels (\ref{13}) allow us 
to compute the energy of  emitted photons. 
The approximate value of the energy for $\mathfrak D\gg1$ and $n\ll\mathfrak D$ in the spectrum  (\ref{13}) is
\begin{eqnarray}\label{13man}
E_n\simeq R_E\left\{-\frac{1}{n^2}-\left(\frac{1}{R_EL_\Lambda}\right)^\frac{2}{3}f_n\right\},~~~
f_n:=4\pi^\frac{8}{3}\left(\frac{1}{4n^4}-\frac{1}{3n^2}+\frac{1}{12}\right),
\end{eqnarray}
where $R_\text{E}=m_\text{e}\alpha^2/2$ is the Rydberg energy. For values $\mathfrak D\gg1$ we can use the non-deformed wave function of the non-perturbed hydrogen atom to calculate the energy shifts of the non-relativistic corrections regarding the presence of the cosmological constant and the corrections of the fine structure, the hyperfine structure and other corrections from quantum field theory such as the Lamb shift and the anomalous magnetic dipole moment of the electron. The result is
\begin{eqnarray}\label{13man1}
E_n\simeq 
R_E\left\{-\frac{1}{n^2}-f_n\left(\frac{1}{R_EL_\Lambda}\right)^\frac{2}{3}
+\frac{A_{nj}}{8}\left(\frac{1}{R_EL_\Lambda}\right)^2
\right\}+{\mathcal O}\left(\frac{1}{n^3}\right).
\end{eqnarray}
The expression of the 
$q-$deformed energy levels (\ref{13man1}) allow us to compute the energy of  emitted photons. 
Furthermore, we can write for the emitted photons a generic expression \begin{multline}\label{13man2}
\frac{1}{\lambda}=
\frac{R_E}{2\pi}\left\{\frac{1}{n_f^2}-\frac{1}{n_i^2}+\delta f_n\left(\frac{1}{R_EL_\Lambda}\right)^\frac{2}{3}
-\frac{1}{8}\delta A_{n,j}\left(\frac{1}{R_EL_\Lambda}\right)^2\right\}+{\mathcal O}\left(\frac{1}{n^3}\right).
\end{multline}
where $n_i,n_f$ are  the  principal  quantum  numbers  of  initial and final states involved in the transition, respectively, $\delta f_n:=f_{n_f}-f_{n_i}$ and $\delta A_{n,j}:=A_{n_f,j_f}-A_{n_i,f_i}$. Note that $R_EL_\Lambda\simeq10^{34}$ and consequently, we can see that the effect of the quantum deformation, $\delta f_n\left(\frac{1}{R_EL_\Lambda}\right)^\frac{2}{3}\simeq\delta f_n\times 10^{-22}$, is {larger} than the effect of curvature, $\frac{\delta A_{nj}}{8}\left(\frac{1}{R_EL_\Lambda}\right)^2\simeq\delta A_{nj}\times 10^{-66}$, but both of them are too small to be measurable with current  spectroscopic  methods. As we will see in the next section, the real impact of the cosmological constant is in the mass of the fundamental particles, number of particles in the universe, and finally in the holographic description of the possible bound states of the atoms.}

{If we neglect the third term in Eq.(\ref{13man2}), which is not relevant to the $q-$deformation, and if we assume large $n$, then we can rewrite it as
\begin{eqnarray}
\frac{1}{\lambda}=R'_\infty\Big(\frac{1}{n_f^2}-\frac{1}{n_i^2}\Big),
\end{eqnarray}
where 
\begin{eqnarray}\label{ryd}
\begin{array}{cc}
R'_\infty:=\frac{m_\text{e}\alpha^2c}{4\pi\hbar}\left\{1-\frac{4\pi^\frac{8}{3}}{3}\left(\frac{2\hbar}{L_\Lambda\alpha^2m_\text{e}c}\right)^\frac{2}{3} \right\}
=R_\infty\left\{1-\frac{2^\frac{4}{3}\pi^2}{3}\left( \frac{1}{R_\infty L_\Lambda}\right)^\frac{2}{3} \right\}.
\end{array}
\end{eqnarray}
is the $q-$deformed Rydberg constant in the SI units. It is pertinent to emphasize herein that Eqs.(\ref{ryd}) and (\ref{15}) (please see next paragraph)  show that the $q-$deformed Rydberg constant is a function of the number of degrees of freedom of electron in the hydrogen atom. The Rydberg constant is one of the most precisely measured physical constants, with a relative standard uncertainty of under two parts in $10^{12}$. The technological (spectroscopic) challenge \cite{pinto1993rydberg} emerges  from the  smallness of $\Lambda$ or equally from a very large value of the cosmic event horizon, $L_\Lambda$, within our currently observed ranges of reach. }

{In more realistic circumstances, the energy density of hydrogen gas could be a source of cosmological dynamics, and we should consider the time of the apparent cosmological horizon that is a boundary hypersurface of an anti-trapped region and has a topology of $\mathbb S^2$. Then we should replace the DeS radius $L_\Lambda$ in Eq. (\ref{ryd}) with the Hubble radius $c/H$. Then, the Hubble parameter is a dynamical variable that satisfies the Friedmann equation
\begin{eqnarray}\label{FF1}
  H^2=H_0^2\left\{\Omega_{0m}(1+z)^3+\Omega_\Lambda\right\},
\end{eqnarray}
where $z$, $H_0$, $\Omega_{0m}=8\pi G\rho_{m}(t_0)/3H_o^2$ and $\Omega_\Lambda=\Lambda c^2/3H_0^2$ are the redshift, the Hubble parameter, the density parameter of the cold matter (dark matter and the hydrogen gas) and the density parameter of the cosmological constant at the present epoch, respectively. In this case, the $q-$deformed Rydberg constant will be a function of the redshift
\begin{eqnarray}
R'_\infty
=R_\infty\left\{1-\frac{2^\frac{4}{3}\pi^2}{3}\left( \frac{H_0}{cR_\infty}\right)^\frac{2}{3}\left(\Omega_{0m}(1+z)^3+\Omega_\Lambda\right)^\frac{1}{3} \right\}.
\end{eqnarray}
As we find, the order of the correction term of the Rydberg constant, $R_\infty=10973731.568160(21) m^{-1}$ at the recombination time, $z=1089$, is in order of $\mathcal O(10^{-9})$ which is out of the range of current measurements. On the other hand, in the radiation dominate area, where the contribution of radiation in the Friedmann equation is given by $\Omega_{0r}(1+z)^4$, the order of correction is $\mathcal O(10^{-5})$, which is in the range of current measurements. Hence, regarding the measurements' current scale, the effects of $q-$deformation are hidden behind the last scattering surface. }

We  close  this  section  mentioning  that, as a result of nilpotent realization \cite{biedenharn1995quantum}, $(J_{\pm}^{(1,2)})^{\mathfrak{D}'}|j_1m_1j_2m_2\rangle=0$,
the Hilbert space of $q-$deformed hydrogen atom is finite-dimensional
\begin{eqnarray}\label{14}
\mathcal H&=\bigoplus_{n=1}^{\mathfrak D'}\mathcal H_n,\cr
\mathcal H_n&=\Big\{|j_1m_1j_2m_2\rangle;\, j_1=j_2=\frac{n-1}{2};m_{i}=-j_{i}(1)j_{i}\Big\},\cr
\mathcal H_{\mathfrak D'}&=\Big\{|j_\text{max}m_1j_\text{max}m_2\rangle;m_{i}=-j_\text{max}(1)j_\text{max}\Big\},
\end{eqnarray}
where $j_\text{max}=\frac{\mathfrak{D}'-1}{2}$ is the azimuthal quantum number of the highest
exited state. Since each $j$ labels a distinct irreducible representation
of $\mathcal U_q(su(2))$ and the number of
$m_i$'s $(m_i= 2j_i + 1)$ is the dimensionality of the representation the dimension of Hilbert space for a $q-$deformed hydrogen atom is
\begin{eqnarray}\label{15}
\text{Dim}\,\,\mathcal H=2^{\sum_{i=1}^{\mathfrak{D}'}n_i^2}\simeq2^{\frac{\mathfrak{D}'^3}{3}}=2^\frac{R_EL_\Lambda}{4\pi}=2^{\frac{m_\text{e}\alpha^2L_\Lambda}{16\pi}}.
\end{eqnarray}

{\subsection{The hydrogen gas in de Sitter space}
To realize the relation of the dimension of that Hilbert space with the entropy
of DeS space, let us now consider a dilute gas of $N$ hydrogen atoms
(as the baryonic matter in late time universe, where  dark energy or the cosmological
constant  dominates) with homogeneous and isotropic
distribution on DeS space. The radial position, $ x$, and the radial
velocity of an atom, $v$, then satisfy the Hubble law $ v= \frac{1}{L_\Lambda}
x$. This suggests that the fluctuations of position and velocity of the atom satisfy
the same equation, $\Delta v= \frac{1}{L_\Lambda}\Delta
x$. The Kinetic energy fluctuations then will $\Delta K=\frac{m_\text{p}}{2}\Delta
v^2=\frac{m_\text{p}}{2L_\Lambda}\Delta  x\Delta v=\frac{1}{4L_\Lambda}$, where $m_\text{p}\simeq m_\text{atom}$ is the
mass of proton and at the last equality we used the uncertainty principle
$\Delta p\Delta x\geqslant1/2$. In the thermodynamical limit $\Delta K/U\simeq1/\sqrt{N}$,
where $U$ is the rest mass of the atom \cite{sivaram1982general}. The above analysis gives
\begin{eqnarray}\label{16}
{N}\simeq (m_\text{p}L_\Lambda)^2 \simeq5.4\times10^{83},
\end{eqnarray}
where we assumed $m_\text{atom}\simeq m_\text{p}$.

Let us summarise some points:

\begin{itemize}
\item First of all, we know that the entropy of a non-relativistic gas {of particles (or dust)} is proportional to the total number of particles, so for hydrogen atom gas, {by considering its components as point-like particles},  we have \cite{rashki2015holography}
\begin{eqnarray}\label{17}
S_\text{gas}\simeq N.
\end{eqnarray}

\item  Furthermore, like the Bekenstein-Hawking entropy of a black hole, the DeS entropy, $S_\text{dS}$,
can be written \cite{gibbons1977cosmological}
\begin{eqnarray}\label{18}
S_\text{dS}=\frac{\pi L_\Lambda^2}{G}=2.88\times10^{122}.
\end{eqnarray}
One can interpret this entropy as the weak holographic principle in which the total number
of degrees of freedom living on the horizon is bounded by one-quarter of the area in Planck units \cite{bianchi2011note,jalalzadeh2017quantum}. 

\item These two entropies (\ref{17}) and (\ref{18}) are not distinct. The total entropy of dilute gas is interrelated to the entropy of DeS space via \cite{rashki2015holography}
\begin{eqnarray}\label{19}
N\simeq S_\text{gas}\simeq S_\text{dS}^\frac{2}{3}=2.02\times10^{81},
\end{eqnarray}
where in the last equality we used the value of $S_\text{dS}$ from (\ref{18}).
The result obtained in (\ref{19}) is consistent \cite{egan2010larger} with the observed value $S_\text{gas}=(9.5 \pm4.5)\times 10^{80}$.

\item Inserting relations (\ref{17}) and (\ref{19}) into (\ref{16}) gives us
\begin{eqnarray}\label{20}
m_\text{p}\simeq\left(\frac{1}{L_\Lambda G}\right)^\frac{1}{3}{\simeq m_\text{Planck}S_\text{dS}^{-\frac{1}{6}}}\simeq\left(\frac{\hbar^2 H_0}{Gc}\right)^\frac{1}{3},
\end{eqnarray}
where $m_\text{Planck}=1/\sqrt{G}$ is the Planck mass and $H_0\simeq c/L_\Lambda$ is the current observed value of the Hubble parameter. The expression in the far right of Eq.(\ref{20})
is the Weinberg formula for the mass of the
nucleon \cite{Weinberg:1972kfs}. Weinberg's relation may then be understood, 
{we speculate},  as the
operational requirement that the mass of the hydrogen atom (or an elementary particle) be such that is not determined solely by local microphysics, but
in the part by the influence of the holographic screen. {As a consequence of Eqs.(\ref{19}) and (\ref{20}), the total mass of hydrogen dust, $M_\text{bulk}$, can be rewritten as
\begin{eqnarray}
M_\text{bulk}\simeq m_\text{atom}N=m_\text{Planck}S_\text{dS}^\frac{1}{2},
\end{eqnarray}
or equivalently 
\begin{eqnarray}
GM_\text{bulk}^2\simeq S_\text{dS}.
\end{eqnarray}
The left-hand side of this relation is the entropy of a black hole the size of the Universe. This shows that the Universe can have no more states than a black hole of the same size.}
\end{itemize}
}

Now, we define the total number of discrete states of all hydrogen atoms in the Universe
by 
\begin{eqnarray}\label{total}
\text{Dim}\,\,\mathcal H_\text{bulk}:=\left(\text{Dim}\,\,\mathcal H\right)^N,
\end{eqnarray}
which leads to 
\begin{eqnarray}\label{21}
\text{Dim}\,\,\mathcal H_\text{bulk}=2^{\frac{m_\text{e}}{m_\text{p}}\alpha^2S_\text{dS}}.
\end{eqnarray}
Given the value ${\alpha^2}\frac{m_\text{e}}{m_\text{p}}\simeq2.9\times10^{-8}$,
it is clear that (\ref{21})  satisfies the holographic bound (\ref{Smolin}).  As it is shown in \cite{bianchi2011note}, the horizon
of DeS is a $2$-dimensional lattice where the number of cells is equal to the DeS entropy
(\ref{18}). Hence, Eq.(\ref{21}) shows that the number of degrees of freedom of all hydrogen atoms in the universe is proportional to the number of cells on the DeS boundary. 
This is congruent with the holographic principle and then the 
holographic entropy bound, which asserts that all natural phenomena within the bulk of a region of
space is {fully realised by the finite set} of degrees of freedom which reside on the boundary, and that this number should not be larger than one binary degree of freedom per Planck area  \cite{bousso1999covariant,hooft1993dimensional}. 

\section{Conclusions and outlook}

\indent 

We conclude by presenting a summary, plus adding a discussion and a brief outlook.

\vspace{0.37cm}

The context that guided our herewith research was that of the holographic entropy bound,  a broad conjecture to apply to all physical systems.  In particular, it was proposed \cite{bousso2004bound} that the total observable entropy in the Universe would be bounded by the inverse of the cosmological constant, including the case of cosmologies dominated by ordinary matter. Such assertion would  constitute a universal law of nature \cite{bousso1999covariant,bousso1999holography}: universes with a positive cosmological constant would be  described by a fundamental theory with only a finite
number of degrees of freedom. This 
is yet to be fully proved and, so far,  it has been broadly tested on cosmological
solutions and suitable gravitational collapsing systems,  within geometrical setups, for
states which have energy eigenvalue
below a threshold  and are localized at space region of specific width. All mentioned reports have confirmed, albeit in restricted configurations,  as remarked within  a significant set of contributions, namely  \cite{bousso2004bound,casini2008relative,bousso2002holographic,bousso2000positive}.

 In more detail, the purpose of our paper was to introduce a discussion on the holographic bound but from another angle. Concretely, we proposed to employ de Sitter holography intertwined with a specific quantization of the hydrogen atom  using the framework of quantum groups. 

Specifically, a concrete   quantum algebra (namely, $\mathcal U_q(so(4))$) was used  to construct a  Hilbert space, whose  retrieved dimension is proportional to $2^{\frac{m_\text{e}\alpha^2L_\Lambda}{16\pi}}$.
We then established that a consequence of the quantum deformation of the hydrogen atom was that the Rydberg constant becomes  dependent on the de Sitter radius, $L_\Lambda$. 
We obtained a finite-dimensional  Hilbert space for the full set of  all hydrogen atoms in
the de Sitter universe. We then showed that the dimension of the latter Hilbert space is $2^{\frac{m_\text{e}}{m_\text{p}}\alpha^2S_\text{dS}}$ and it satisfies the 
holographic entropy bound.
It is well-known that to formulate quantum electrodynamics, we just need two dimensionless constants: the first one is the fine structure constant, $\alpha$, and the second one is the ratio of the electron mass to the proton mass $\beta=\frac{m_\text{e}}{m_\text{p}}$ \cite{Barrow1}. Apart from numerical factors like the atomic number, $Z$, or integral quantum numbers, the whole physical properties of atoms, molecules, and solids can be determined as functions of $\alpha$ and $\beta$ \cite{Barrow2}. Equation (\ref{21}) shows that these two parameters also play a crucial role in the holographic bound of the hydrogen atom gas.
Furthermore, we 
also expressed  that the mass of a hydrogen atom $m_\text{atom}$ and the total number of atoms inside the  cosmic  event  horizon, $N$, are related (through simple expressions that the holography bound conjecture endorses) to  the de Sitter entropy, $S_\text{dS}$ and the Planck mass, $m_\text{Planck}$, by $m_\text{atom}\simeq m_\text{Planck}S_\text{dS}^{-\frac{1}{6}}$, and $N\simeq S_\text{dS}^\frac{2}{3}$.

Although we used a simple model, we are confident it can be extended to the case of radiation  or even more elaborated, a spin-1 field theory description within the framework we used, even if restricted to a de Sitter space. Perhaps bolder, gravitational degrees of freedom could eventually be considered  within quantum groups and constructing the Hilbert space, hopefully finite.   Thus, we trust the features of quantum deformation (within quantum groups) may be considered as reliable complementary  tool to explore holography, herein  brought in an interesting  intertwined manner.

\section*{Acknowledgments}
This research work was supported by Grants No. 
UID-B/00212/2020, UID-P/00212/2020 and COST Action CA18108 (Quantum gravity phenomenology in the multi-messenger approach).

\section*{References}
\bibliographystyle{iopart-num}
\bibliography{Atom.bib}

\providecommand{\newblock}{}
\begin{thebibliography}{10}
\expandafter\ifx\csname url\endcsname\relax
  \def\url#1{{\tt #1}}\fi
\expandafter\ifx\csname urlprefix\endcsname\relax\def\urlprefix{URL }\fi
\providecommand{\eprint}[2][]{\url{#2}}

\bibitem{bekenstein1981universal}
Bekenstein J~D 1981 {\em Phys. Rev. D\/} {\bf 23} 287

\bibitem{bekenstein1973black}
Bekenstein J~D 1973 {\em Phys. Rev. D\/} {\bf 7} 2333

\bibitem{bekenstein1974generalized}
Bekenstein J~D 1974 {\em Phys. Rev. D\/} {\bf 9} 3292

\bibitem{hawking1975particle}
Hawking S~W 1975 {\em Commun. Math. Phys.\/} {\bf 43} 199--220

\bibitem{bousso1999covariant}
Bousso R 1999 {\em J. High Energy Phys.\/} {\bf 1999} 004

\bibitem{2020arXiv200912530B}
{Buoninfante} L, {Gaetano Luciano} G, {Petruzziello} L and {Scardigli} F 2020
  {\em arXiv e-prints\/} arXiv:2009.12530 (\textit{Preprint}
  \eprint{2009.12530})

\bibitem{10.2307/26060403}
Bekenstein J~D 2003 {\em Scientific American\/} {\bf 289} 58--65

\bibitem{tHooft:1993dmi}
{'t Hooft} G 1993 {\em arXiv e-prints\/} gr-qc/9310026 (\textit{Preprint}
  \eprint{gr-qc/9310026})

\bibitem{fischler1998holography}
{Fischler} W and {Susskind} L 1998 {\em arXiv e-prints\/} hep-th/9806039
  (\textit{Preprint} \eprint{hep-th/9806039})

\bibitem{susskind1995world}
Susskind L 1995 {\em J. Math. Phys.\/} {\bf 36} 6377--6396

\bibitem{smolin2001strong}
Smolin L 2001 {\em Nucl. Phys. B\/} {\bf 601} 209--247

\bibitem{bousso2004bound}
Bousso R 2004 {\em J. High Energy Phys.\/} {\bf 2004} 025

\bibitem{casini2008relative}
Casini H 2008 {\em Class. Quantum Grav.\/} {\bf 25} 205021

\bibitem{bousso2002holographic}
Bousso R 2002 {\em Rev. Mod. Phys.\/} {\bf 74} 825

\bibitem{bousso2000positive}
Bousso R 2000 {\em J. High Energy Phys.\/} {\bf 2000} 038

\bibitem{2008PhRvD..78f4047A}
{Ashtekar} A and {Wilson-Ewing} E 2008 {\em Phys. Rev. D\/} {\bf 78} 064047
  (\textit{Preprint} \eprint{0805.3511})

\bibitem{2000CQGra..17L..83B}
{Bak} D and {Rey} S~J 2000 {\em Class. and Quantum Grav.\/} {\bf 17} L83--L89
  (\textit{Preprint} \eprint{hep-th/9902173})

\bibitem{bousso1999holography}
Bousso R 1999 {\em J. High Energy Phys.\/} {\bf 1999} 028

\bibitem{moradi2010hydrogen}
Moradi S and Aboualizadeh E 2010 {\em Gen. Relativ. Gravit.\/} {\bf 42}
  435--442

\bibitem{parker1980one}
Parker L 1980 {\em Phys. Rev. Lett.\/} {\bf 44} 1559

\bibitem{parker1982gravitational}
Parker L and Pimentel L~O 1982 {\em Phys. Rev. D\/} {\bf 25} 3180

\bibitem{chaichian1996introduction}
Chaichian M and Demichev A~P 1996 {\em Introduction to quantum groups\/} (World
  Scientific)

\bibitem{kupriyanov2013hydrogen}
Kupriyanov V~G 2013 {\em J. Phys. A: Math. Theor.\/} {\bf 46} 245303

\bibitem{chaichian2004non}
Chaichian M, Sheikh-Jabbari M~M and Tureanu A 2004 {\em Eur. Phys. J. C\/} {\bf
  36} 251--252

\bibitem{Kustaanheimo:419610}
Kustaanheimo P and Stiefel E 1965 {\em J. Reine Angew. Math.\/} {\bf 218}
  204--219

\bibitem{boiteux1973three}
Boiteux M 1973 {\em Physica\/} {\bf 65} 381--395

\bibitem{kibler1991q}
Kibler M and N{\'e}gadi T 1991 {\em J. Phys. A: Math. Gen.\/} {\bf 24} 5283

\bibitem{song1992quantum}
Song Z~C and Liao L 1992 {\em J. Phys. A: Math. Gen.\/} {\bf 25} 623

\bibitem{gora1992two}
Gora J 1992 {\em J. Phys. A: Math. Gen.\/} {\bf 25} L1281

\bibitem{yang1993energy}
Yang Q~G and Xu B~W 1993 {\em J. Phys. A: Math. Gen.\/} {\bf 26} L365

\bibitem{castro2015physics}
Castro P~G and Kullock R 2015 {\em Theo. Math. Phys.\/} {\bf 185} 1678--1684

\bibitem{kulish1983quantum}
Kulish P~P and Reshetikhin N~Y 1983 {\em J. Soviet Math.\/} {\bf 23} 2435--2441

\bibitem{sklyanin1982some}
Sklyanin E~K 1982 {\em Funct. Anal. its Appl.\/} {\bf 16} 263--270

\bibitem{faddeev2016liouville}
Faddeev L~D and Takhtajan L~A 2016 Liouville model on the lattice {\em Fifty
  Years of Mathematical Physics: Selected Works of Ludwig Faddeev\/} (World
  Scientific) pp 159--172

\bibitem{chari1995guide}
Chari V, Pressley A~N {\em et~al.\/} 1995 {\em A guide to quantum groups\/}
  (Cambridge university press)

\bibitem{oh1992conformal}
Oh C and Singh K 1992 {\em J. Phys. A: Math. Gen.\/} {\bf 25} L149

\bibitem{kauffman2007q}
Kauffman L~H and Lomonaco~Jr S~J 2007 {\em J. Knot Theory and Its
  Ramifications\/} {\bf 16} 267--332

\bibitem{foda1994vertex}
Foda O, Jimbo M, Miwa T, Miki K and Nakayashiki A 1994 {\em J. Math. Phys.\/}
  {\bf 35} 13--46

\bibitem{nayak2008non}
Nayak C, Simon S~H, Stern A, Freedman M and Sarma S~D 2008 {\em Rev. Mod.
  Phys.\/} {\bf 80} 1083

\bibitem{chang1991q}
Chang Z, Guo H~Y and Yan H 1991 {\em Phys. Lett. A\/} {\bf 156} 192--196

\bibitem{smolin1995linking}
Smolin L 1995 {\em J. Math. Phys.\/} {\bf 36} 6417--6455

\bibitem{major1996quantum}
Major S and Smolin L 1996 {\em Nucl. Phys. B\/} {\bf 473} 267--290

\bibitem{dupuis2013quantum}
Dupuis M and Girelli F 2013 {\em Phys. Rev. D\/} {\bf 87} 121502

\bibitem{dittrich2014quantum}
Dittrich B, Martin-Benito M and Steinhaus S 2014 {\em Phys. Rev. D\/} {\bf 90}
  024058

\bibitem{rovelli2015compact}
Rovelli C and Vidotto F 2015 {\em Phys. Rev. D\/} {\bf 91} 084037

\bibitem{biedenharn1995quantum}
Biedenharn L~C and Lohe M~A 1995 {\em Quantum group symmetry and $q$-tensor
  algebras\/} (World Scientific)

\bibitem{adams2012algebraic}
Adams B~G 2012 {\em Algebraic approach to simple quantum systems: with
  applications to perturbation theory\/} (Springer Science \& Business Media)

\bibitem{pauli1926wasserstoffspektrum}
Pauli W 1926 {\em Zeitschrift f{\"u}r Physik\/} {\bf 36} 336--363

\bibitem{pinto1993rydberg}
Pinto F 1993 {\em Phys. Rev. Lett.\/} {\bf 70} 3839

\bibitem{sivaram1982general}
Sivaram C 1982 {\em Astrophys. Space Sci.\/} {\bf 86} 501--504

\bibitem{rashki2015holography}
Rashki M and Jalalzadeh S 2015 {\em Phys. Rev. D\/} {\bf 91} 023501

\bibitem{gibbons1977cosmological}
Gibbons G~W and Hawking S~W 1977 {\em Phys. Rev. D\/} {\bf 15} 2738

\bibitem{bianchi2011note}
Bianchi E and Rovelli C 2011 {\em Phys. Rev. D\/} {\bf 84} 027502

\bibitem{jalalzadeh2017quantum}
Jalalzadeh S, Capistrano A~J~S and Moniz P~V 2017 {\em Phys. Dark Universe\/}
  {\bf 18} 55--66 (\textit{Preprint} \eprint{1709.09923})

\bibitem{egan2010larger}
Egan C~A and Lineweaver C~H 2010 {\em The Astrophysical Journal\/} {\bf 710}
  1825

\bibitem{Weinberg:1972kfs}
Weinberg S 1972 {\em {Gravitation and Cosmology}: {Principles and Applications
  of the General Theory of Relativity}\/} (New York: John Wiley and Sons)

\bibitem{hooft1993dimensional}
{'t Hooft} G 1993 {\em arXiv e-prints\/} gr-qc/9310026 (\textit{Preprint}
  \eprint{gr-qc/9310026})

\bibitem{Barrow1}
{Barrow} J~D 2002 {\em {The constants of nature : from alpha to omega}\/}
  (London: Jonathan Cape)

\bibitem{Barrow2}
{Barrow} J~D and {Tipler} F~J 1986 {\em {The anthropic cosmological
  principle.}\/} (Oxford: Oxford UP)

\end{thebibliography}

\end{document}